\magnification=\magstep 1
\tolerance 500
\rightline{IASSNS-HEP-96/99}
\rightline{TAUP 2379-96}
\vskip 3 true cm
\centerline{\bf Equilibrium Statistical Ensembles}
\centerline{\bf and}
\centerline{\bf Structure of the Entropy Functional}
\centerline{\bf in}
\centerline{\bf Generalized Quantum Dynamics}\footnote{$^*$}{ Presented
 by the second author at Quantum Structures '96, Berlin
July 29-Aug. 3, 1996.}
\vskip 1 true cm
\centerline{Stephen L.  Adler
and L.P. Horwitz\footnote{$^{**}$}{On Sabbatical leave from
School of Physics, Raymond and Beverly Sackler Faculty of Exact
Sciences, Tel Aviv University, Ramat Aviv 67798, Israel, and
 Department of Physics,
Bar Ilan University, Ramat Gan 52900, Israel.}}
\smallskip
\centerline{School of Natural Sciences}
\centerline{ Institute for
Advanced Study, Princeton, N.J. 08540, U.S.A.}

\vskip 2 true cm
\noindent
{\it Abstract \/}: We review here the microcanonical and canonical
ensembles constructed on an underlying generalized quantum dynamics
and the algebraic properties of the conserved quantities.  We discuss
the structure imposed on the microcanonical entropy by the equilibrium
conditions.

\vfill
\eject
\noindent
{\bf 1. Introduction}
\par In this paper we review briefly the generalized quantum
dynamics$^{1,2}$ constructed on a phase space of local noncommuting
fields. We show that the equilibrium conditions on the microcanonical
entropy imply that the
system decomposes thermodynamically to a
sequence of adiabatically independent subsystems, each with its own
temperature. There is an equipartition theorem for the phase space
variables of the system generated by the  linear combination
of conserved quantities associated with each of these independent
thermodynamic modes.
\par We start with a review of our basic framework.
 Generalized quantum dynamics$^{1,2}$ is an analytic mechanics on
a symplectic set of operator valued variables, forming an operator
valued phase space $\cal S$.  These variables are defined as the set of
linear transformations\footnote{$^\dagger$}{In general,
 local (noncommuting) quantum
fields.}
 on an underlying real, complex, or quaternionic
Hilbert space (Hilbert module), for which the postulates of a real,
complex, or quaternionic quantum mechanics are satisfied$^{2-6}$.  The
dynamical (generalized Heisenberg) evolution,
or flow, of this phase space is
generated by the total trace Hamiltonian ${\bf H} = {\bf Tr} H$, where
for any operator ${\rm O}$ we have
$$ \eqalign{
{\bf O}\equiv{\bf Tr} {\rm O} &\equiv {\rm Re Tr} (-1)^F {\rm O} \cr
                              &= {\rm Re} \sum_n \langle n \vert (-1)^F
{\rm O} \vert n \rangle,\cr} \eqno(1.1)$$
$H$ is a function of the operators $\{ q_r(t) \}, \{p_r(t) \},\ \ {\rm
r}= 1,2, \dots,N$ (realized as a sum of monomials, or a limit of a
sequence of such sums; in the general case of local noncommuting
fields, the index $r$ contains continuous variables),
and $(-1)^F$ is a grading operator with
eigenvalue $1(-1)$ for states in the boson (fermion) sector of the
Hilbert space. Operators are called bosonic or fermionic in type if they
commute or anticommute, respectively, with $(-1)^F$; for each $r$, $p_r$
and $q_r$ are of the same type.
\par The variation of a total trace functional with respect to some
operator is defined with the help of the cyclic property of
the ${\bf Tr}$ operation.  The variation of any monomial ${\rm O}$
consists of terms of the form ${\rm O}_L \delta x_r {\rm O}_R$, for
$x_r$ one of the $\{q_r \}, \{p_r\}$,
which, under the ${\bf Tr}$ operation, can be brought to the form
$$ \delta {\bf O} = \delta {\bf Tr}{\rm O} = \pm {\bf Tr}
{\rm O}_R {\rm O}_L \delta x_r, $$
so that sums and limits of sums of such monomials permit the
construction of
$$ \delta {\bf O} = {\bf Tr} \sum_r {\delta {\bf O} \over
\delta x_r } \delta x_r,\eqno(1.2) $$
uniquely defining ${\delta {\bf O}}/{\delta x_r}$.
\par Assuming the existence of a total trace Lagrangian$^{1,2}$ ${\bf
L} = {\bf L} (\{q_r \}, \{ {\dot q}_r \} )$, the variation of the
total trace action
$$ {\bf S} = \int_{-\infty}^\infty  {\bf L} (\{q_r \}, \{ {\dot q}_r
\}) dt   \eqno(1.3) $$
results in the operator Euler-Lagrange equations
$$  {\delta {\bf L} \over {\delta q_r}} - {d \over {dt}} {{\delta {\bf
L}} \over {\delta \dot q_r} } = 0 . \eqno(1.4)$$
As in classical mechanics, the total trace Hamiltonian is defined as a
Legendre transform,
$$ {\bf H} = {\bf Tr} \sum_r p_r \dot q_r - {\bf L},\eqno(1.5) $$
where
$$ p_r = { {\delta {\bf L}} \over {\delta \dot q_r}}.  \eqno(1.6)$$
 It then follows from $(1.4)$ that
$${ {\delta {\bf H}} \over {\delta q_r} } = -\dot p_r  \qquad { {\delta
{\bf H}} \over {\delta p_r}} = \epsilon_r \dot q_r,\eqno(1.7) $$
where $\epsilon_r = 1(-1)$ according to whether $p_r, q_r$ are of
bosonic (fermionic) type.
\par Defining the generalized Poisson bracket
$$ \{{\bf A}, {\bf B} \} = {\bf Tr} \sum_r \epsilon_r \left( {\delta
{\bf A} \over \delta q_r} {\delta {\bf B}\over \delta p_r} -
{\delta {\bf B} \over \delta q_r} {\delta {\bf A}\over \delta
p_r}\right), \eqno(1.8a) $$
one sees that
$$ {d{\bf A} \over dt } = {\partial {\bf A} \over \partial t} +
\{{\bf A}, {\bf H} \}. \eqno(1.8b) $$
Conversely, if we define
$$ {\bf x}_s(\eta) = {\bf Tr} (\eta x_s) , \eqno(1.9a)$$
for $\eta$ an arbitrary, constant operator (of the same type as $x_s$,
which denotes here $q_s$ or $p_s$), then
 $$  {d{\bf x}_s(\eta)\over dt} =
 {\bf Tr} \sum_r \epsilon_r \left( {\delta
{\bf x}_s(\eta) \over \delta q_r} {\delta {\bf H}\over \delta p_r} -
{\delta {\bf H} \over \delta q_r} {\delta {\bf x}_s(\eta)\over \delta
p_r}\right), \eqno(1.9b)$$
and comparing the coefficients of $\eta$
on both sides, one obtains the Hamilton equations $(1.7)$ as a
consequence of the Poisson bracket relation $(1.8b)$.
\par The Jacobi identity is satisfied by the Poisson bracket
$(1.8a)$,$^7$ and hence the total trace functionals have many of the
properties of the corresponding quantities in classical mechanics.$^8$
In particular, canonical transformations take the form
$$\delta {\bf x}_s(\eta)=\{ {\bf x}_s(\eta), {\bf G} \}, \eqno(1.10a)$$
which implies that
$$\delta p_r= -{\delta {\bf G} \over \delta q_r}~,~~~
\delta q_r=\epsilon_r  {\delta {\bf G} \over \delta p_r}~,\eqno(1.10b)$$
with the generator ${\bf G}$ any total trace functional constructed from
the operator phase space variables.
 Time evolution then corresponds to the
special case ${\bf G}={\bf H} dt$.
\par It has recently been shown by Adler and Millard$^9$ that a
canonical ensemble can be constructed on the
phase space ${\cal S}$, reflecting
the equilibrium properties of a system of many degrees of freedom.
Since the operator
$$ \eqalign{{\tilde C} &= \sum_r (\epsilon_r q_r p_r - p_r q_r) \cr
&=\sum_{r,B} [q_r,p_r] - \sum_{r,F} \{q_r,p_r\},\cr}
\eqno(1.11)$$
where the sums are over bosonic and fermionic pairs, respectively, is
conserved under the evolution $(1.7)$ induced by the total trace
Hamiltonian, the canonical ensemble must be constructed taking this
constraint into account. This is done by including in the
canonical exponent the conserved
quantity ${\bf Tr}{\tilde \lambda}{\tilde C}$, for some given constant
anti-hermitian operator ${\tilde \lambda}$.
\par In the general case, in the presence of the fermionic sector,
the graded trace of the Hamiltonian is not bounded from below, and the
partition function may be divergent. When the equations of motion
induced by the  Lagrangian ${\bf L}$ coincide with those induced by
the ungraded total trace of the same Lagrangian,
$$ {\hat {\bf L}} = {\rm Re Tr} L, $$
without the factor $(-1)^F$, the corresponding ungraded total trace
Hamiltonian ${\hat {\bf H}}$ is conserved; it may therefore be
included as a constraint functional in the canonical ensemble, along
with the new conserved quantity ${\hat{\bf Tr}}
{\hat{\tilde \lambda}}{\hat{\tilde C}}$ (see Appendices 0 and C of ref.
9),
 where
$$ \eqalign{{\hat{\tilde C}} &= \sum_r [q_r, p_r]\cr
 &= \sum_{r,B} [q_r,p_r] + \sum_{r,F} [q_r,p_r] .\cr}
\eqno(1.12) $$
   It was argued that the Ward identities derived from the
canonical ensemble imply that ${\hat{\tilde \lambda}}$ and
${\tilde \lambda}$ are functionally related, so that they may be
diagonalized in the same basis (Appendix F of ref. 9).
It was then shown that, since the ensemble averages depend only on
${\tilde \lambda}$ and $(-1)^F$, the ensemble average of any operator
must commute with these operators.  Since the ensemble averaged  operator
 $\langle {\tilde C}\rangle_{AV}$ is
anti-self-adjoint, if one furthermore assumes it is completely
degenerate (with eigenvalue $i_{eff} \hbar$), the ensemble average of
the theory then reduces to the usual complex quantum field theory.
\par As discussed in detail in ref.10, the phase space volume associated
with the microcanonical ensemble can be written as
$$ \eqalign{\Gamma( E, {\hat E}, \tilde \nu,
{\hat {\tilde \nu}}) &= \int d\mu\,\delta( E-{\bf
H})\,\delta({\hat E}- {\hat{\bf H}})\cr
& \prod_{n\leq m,A}\, \delta(\nu_{nm}^A
- \langle n \vert (-1)^F {\tilde C} \vert m \rangle^A)\, \delta({\hat
\nu}_{nm}^A - \langle n \vert {\hat{\tilde C}} \vert m
\rangle^A ),\cr}  \eqno(1.13)$$
where we have taken into account the possible algebraic structure
of the matrix elements of the operators with the  index $A$, which
takes the values $0,\ 1$ for the complex Hilbert space, $0,\ 1,\ 2,
\ 3$ for the quaternionic Hilbert space, and just the value $0$ for
real Hilbert space. The invariant phase space measure is defined by
$$ \eqalign{ d\mu &= \prod_A d\mu^A, \cr
      d\mu^A &\equiv \prod_{r,m,n} d(x_r)^A_{mn},\cr } \eqno(1.14)$$
where redundant factors are omitted according to adjointness
conditions.  We have, furthermore, used the
 abbreviations $\tilde \nu \equiv \{\nu_{nm}^A \}$ and
${\hat {\tilde \nu}} \equiv \{{\hat \nu}_{nm}^A \}$.
The entropy associated with this ensemble is  given by
$$ S_{mic}(E, {\hat E}, \tilde {\nu}, {\hat {\tilde \nu}}) =
\ln \, \Gamma(E, {\hat E},\tilde \nu,{\hat{\tilde \nu}}).\eqno(1.15)$$
\par It was argued in ref. 10 that a large system can be decomposed into
a part within a certain (large) region of the measure space, which we
denote as $b$, corresponding to what we shall consider as a {\it
bath}, in the sense of statistical mechanics, and another (small)
part which we shall denote as $s$, corresponding to what we shall
consider as a {\it subsystem}.  It was then argued$^{10}$ that the
phase space volume can be well approximated by
$$ \Gamma(E, {\hat E}, \tilde \nu, {\hat {\tilde \nu}})
= \int dE_s d{\hat E}_s
(d\nu^s)(d{\hat
\nu}^s) \, \Gamma_b (E-E_s, {\hat E}-{\hat E}_s,\tilde \nu - \tilde \nu_s,
{\hat {\tilde \nu}} - {\hat {\tilde \nu}}_s)\,
\Gamma_s(E_s, {\hat E}_s,\tilde \nu_s, {\hat {\tilde \nu}}_s).
 \eqno(1.16)$$
\par Defining the variables
$$ \xi = \{ \xi_i \} \equiv \{E, {\hat E}, {\tilde \nu},
 {\hat {\tilde \nu}}\},
 \eqno(1.17)$$
it was shown$^{10}$ that the equilibrium conditions which follow from the
assumption that there is a maximum in the integrand of Eq. $(1.16)$
(which dominates the integral in the limit of a large number of degrees
of freedom) result in the set of equalities
$$ { 1 \over \Gamma_s (\xi)} {\partial \Gamma_s  \over
\partial \xi_i}  (\xi) |_{\bar \xi} = { 1 \over \Gamma_b(\Xi - \xi)}
{\partial  \Gamma_b\over \partial \Xi_i }(\Xi -
\xi)|_{\bar \xi},
\eqno(1.18)$$
where $\Xi$ corresponds to the total quantities belonging to the
full system. It was then shown that the canonical ensemble obtained
by Adler and Millard$^{1,2}$,
 $$ \rho = Z^{-1} \exp- \{ \tau {\bf H} +
{\hat \tau} {\hat {\bf H}} + {\bf Tr} {\tilde \lambda} {\tilde
C} + {\hat {\bf Tr}} {\hat {\tilde  \lambda}} {\hat {\tilde C}}
\}, \eqno(1.19)$$
where $$ Z = \int d\mu\, \exp- \{ \tau {\bf H} +
{\hat \tau} {\hat {\bf H}} + {\bf Tr} {\tilde \lambda} {\tilde
C} + {\hat {\bf Tr}} {\hat {\tilde  \lambda}} {\hat {\tilde C}}
\}, \eqno(1.20)$$
follows in a straightforward way. The quantities $\tau,\  {\hat \tau}$
and the matrices (real, complex, or quaternionic) $\lambda,\ {\tilde
\lambda}$ are the equilibrium parameters defined by the values of
the members of $(1.18)$ for each of the $\xi$'s$^{10}$; they therefore
correspond to {\it temperatures} precisely as they emerge in
conventional statistical mechanics.  We remark that Ingarden$^{11}$
has studied a similar generalization of temperature in the framework
of the statistical mechanics associated with problems of optical
pumping (in the diagonal form which we shall discuss in the next
section).
\par Replacing the operators and
trace functionals in $(1.20)$ by integrals over $\delta$-functions,
 the partition function can be rewritten as$^{10}$
$$   Z= \int dE d{\hat E} (d\nu)(d{\hat \nu}) e^{S_{mic}(E,
{\hat E}, {\tilde \nu},{ \hat{\tilde \nu}})} \exp-\{\tau  E +
{\hat \tau} \hat E + {\bf Tr} {\tilde \lambda} {\tilde \nu} +
{\hat {\bf Tr}} {\hat {\tilde  \lambda}} {\hat {\tilde \nu}}
\}. \eqno(1.21)  $$
\par By studying the dispersions of the variables in the
canonical ensemble, it was found$^{10}$ that the second derivative
matrix of the microcanonical entropy is negative definite, i.e.,
that
$$\bigl({\partial^2 S_{mic}\over \partial \xi_i \partial \xi_j}\bigr)
 \leq 0  \eqno(1.22)$$
\par In the following we use the fact that this matrix is
real symmetric to diagonalize it, and in this way
to construct a set of dynamical generators over which the total entropy
decomposes in a neighborhood $C_0$ of the maximum entropy point.
\bigskip
\noindent
{\bf 2. Diagonal form of the second variation of the entropy}
\smallskip
\par  The negative definite matrix Eq. $(1.22)$
$$ D_{ij} = {\partial^2 S_{mic}\over \partial
\xi_i \partial \xi_j}  \eqno(2.1)$$
is symmetric and can therefore be diagonalized by an orthogonal
transformation.  Let $a_{ij}$ (orthogonal) be such that, in the
neighborhood $C_0$,
$$ \sum_{ij} a_{ki} a_{\ell j} D_{ij} =
\delta_{kl} d_\ell (\xi), \eqno(2.2)$$
where the elements $d_\ell(\xi)$ on the right hand side
are the negative
eigenvalues.  Now, let us define, using these constant coefficients,
$$e_k = \sum_i a_{ki}\xi_i, \eqno(2.3a)$$
and hence
$$ \xi_i = \sum_k a_{ki} e_k .\eqno(2.3b)$$
It then follows that, in $C_0$,
$$ \eqalign{{\partial^2 S \over
{\partial e_k \partial e_\ell}}
&= \sum_{ij} a_{ki} a_{\ell j} { \partial^2 S \over {\partial \xi_i
\partial \xi_j}} \cr &= \sum_{ij} a_{ki} a_{\ell j} D_{ij} =
\delta_{k\ell} d_\ell (\xi).\cr} \eqno(2.4)$$
Since the crossed derivatives of $S$ vanish, $S$ must be a sum
of functions that depend on each of the $\{e_k \}$ separately, i.e.,
$$ S = \sum_k S_k (e_k). \eqno(2.5)$$
The entropy is therefore additive (in $C_0$) over
diagonal ``thermodynamic
modes''.
\par The equilibrium parameters defined in the previous section,
$$ \chi_j = {\partial S \over \partial \xi_j} = \{\tau,\ {\hat{\tau}},
\ \lambda,\  {\hat{\lambda}}\}, \eqno(2.6)$$
may be transformed in the same way, i.e.,
$$ \eqalign{\sum_j a_{kj} \chi_j &= \sum_j a_{kj} {\partial \over
\partial \xi_j} S = {\partial S \over \partial e_k} \cr
&= {\partial S_k(e_k) \over \partial e_k} \equiv {1 \over T_k}, \cr}
\eqno(2.7)$$
giving the diagonal temperatures (of the type
considered by Ingarden$^{11}$).
\par  We remark that, according to $(2.2)$
and $(2.4)$,
$$ {\partial^2 S \over \partial e_k^2} = d_k <0, \eqno(2.8)$$
so that the ``specific heats'', entering as
$$ {\partial \over \partial e_k} {1 \over T_k} = - {1 \over
T_k^2} {dT_k \over de_k} = - {1 \over T_k^2} {1 \over C_k},
\eqno(2.9)$$
are positive, and by $(2.7)-(2.9)$ are given by
 $$ C_k = - {1 \over T_k^2}d_k. \eqno(2.10)$$
\bigskip
\noindent
{\bf 3. Equipartition}
\smallskip
\par We now consider linear combinations of the dynamical
quantities $${\cal H}_i= \{ {\bf H}, {\hat{\bf H}}, {\tilde C},
{\hat{\tilde C}} \}$$ of the same form as the
 linear combinations of the parameters $\{\xi_i \}$ which are their
equilibrium values,
$$ \varepsilon_k = \sum_{i} a_{ki} {\cal H}_i, \eqno(3.1)$$
the effective ``energies'' associated with the thermodynamic modes.
Since the determinant of the matrix $a$ is unity, the microcanonical
phase space integral $(1.13)$ can be written as
$$\Gamma(\xi) = \int d\mu  \prod_k \delta(e_k - \varepsilon_k);
\eqno(3.2)$$
since, however, as we have shown in Section 2,
$$ \eqalign{ \ln \Gamma(\xi) &= S(e_1, e_2, \dots ) \cr
       &= \sum_k S_k (e_k) ,\cr}\eqno(3.3)$$
it follows that the phase space volume factorizes on the diagonal
parameters
$$\eqalign{\Gamma(\xi) & = e^{\sum_k S_k(e_k)}
= \prod_k e^{S_k(e_k)}  \cr
&= \Gamma(e_1,e_2, \dots ) \equiv \prod_k \Gamma_k(e_k). \cr}
\eqno(3.4)$$
One can show that the free energy also becomes additive$^{12}$.
\par Let us now consider the microcanonical average
$$ \bigl\langle q_r {\delta \varepsilon_k
\over \delta q_s} \bigr\rangle
= {1 \over \Gamma(e_1, e_2, \dots )} \int d\mu \prod_\ell
 \delta (e_\ell-\varepsilon_\ell) q_r {\delta \varepsilon_k
\over \delta q_s},\eqno(3.5)$$
where $\{q_r\}$ are the canonical coordinates (fields) of the
phase space.  We now write the right hand side of $(3.5)$ identically
as
$$ {1 \over \Gamma(e_1, e_2, \dots)}\prod_\ell {\partial \over
\partial e_\ell} \int_{\{\varepsilon_j < e_j\}}
d\mu \ q_r {\delta \over \delta q_s}(\varepsilon_k - e_k), $$
replacing the $\delta$-functions by derivatives of the parameters
of boundary step functions; adding the constant $e_k$ does
not affect the result. Integrating by parts in the integration over
phase space, we obtain
$$ \bigl\langle q_r {\delta \varepsilon_k \over \delta q_s}
\bigr\rangle = {1 \over \Gamma(e_1, e_2, \dots )} \prod_\ell
{\partial \over \partial e_\ell} \int_{\varepsilon_j < e_j}
\bigl\{ {\delta \over \delta q_s} q_r (\varepsilon_k - e_k)
- \delta_{rs} (\varepsilon_k - e_k) \bigr\}.\eqno(3.6)$$
The first term vanishes on the boundary, and we therefore have
$$ \bigl\langle q_r {\delta \varepsilon_k \over \delta q_s} \bigr
\rangle = -{\delta_{rs} \over \Gamma(e_1, e_2, \dots )}
\prod_\ell {\partial \over \partial e_\ell}
\int_{\{\varepsilon_j < e_j\}} d\mu (\varepsilon_k - e_k). \eqno(3.7)$$
\par The derivative with respect to $e_k$ in the product of
derivatives vanishes when it differentiates the upper bound; its
contribution is only from the integrand, resulting in a factor $-1$.
The other derivatives act only on the upper limits.  The product
then results in the restricted measure
$$ \int_{\varepsilon_k < e_k} d\mu (\varepsilon_\ell = e_\ell \ \
 \forall \ \
\ell \neq k) \equiv \Sigma_k, \eqno(3.8)$$
which can be rewritten as
$$ \Sigma_k = \int_{\{\varepsilon_j < e_j\}} d\mu \prod_{\ell \neq k}
\delta ( \varepsilon_\ell - e_\ell). \eqno(3.9)$$
 According to $(3.2)$,
$$ {\partial \Sigma_k \over \partial e_k} = \Gamma(e_1, e_2, \dots).
\eqno(3.10)$$
\par We therefore have
$$\bigl\langle q_r {\delta \varepsilon_k \over \delta q_s}
\bigr\rangle
= {\delta_{rs} \over \Gamma(e_1, e_2, \dots) } \Sigma_k.\eqno(3.11)$$
We now use the factorization of $\Gamma(e_1, e_2, \dots)$
in $C_0$ to derive a
relation between $\Sigma_k$ and the additive entropies.  In the limit of
a large number of degrees of freedom, the leading edge of the integral
defining $\Sigma_k$ dominates the integral$^{13}$, so we may formally
extrapolate, as a model, the quadratic form (and associated
factorization) valid in $C_0$. From $(3.4)$,
and $(3.10)$ it then follows that
$$ {\partial \Sigma_k \over \partial e_k} =
 \Gamma_k (e_k) \prod_{\ell \neq k} \Gamma_\ell(e_\ell). \eqno(3.12)$$
One may integrate this equation to obtain
$$  \Sigma_k =
 \int^{e_k} \ \Gamma_k (e_k')de_k' \prod_{\ell \neq k}
 \Gamma_\ell(e_\ell) + G(e_\ell, \ell \neq k). \eqno(3.13)$$
Since $\varepsilon_k$ cannot be $-\infty$ (the functional
${\hat{\bf H}}$ is contained linearly and its positive values
are assumed to dominate for large values of the phase space variables),
the first term on the right hand side of $(3.13)$, along with
$\Sigma_k$ must vanish as $e_k \rightarrow -\infty$, and hence
G must be zero.
\par We therefore obtain
$$ \Sigma_k = \int^{e_k} de_k' \ e^{S_k(e_k')} \prod_{\ell \neq k}
e^{S_\ell(e_\ell)}, \eqno(3.14)$$
so that
$$\eqalign {{\Sigma_k \over \Gamma (e_1, e_2, \dots)} &=
{\int^{e_k} de_k' e^{S_k(e_k')} \over e^{S_k(e_k) }}\cr
&= {1 \over {d\over de_k} \ln \int^{e_k} de_k' \  e^{S_k(e_k')}}. \cr}
\eqno(3.15) $$
With the leading approximation for a large number of degrees of
freedom$^{13}$,
$$ \ln \int^{e_k} de_k' \ e^{S_k(e_k')} \sim \ln e^{S_k(e_k)}, $$
we conclude that
$$ \bigl\langle q_r {\delta \varepsilon_k \over \delta q_s} \bigr\rangle
= -\delta_{rs} T_k. \eqno(3.16)$$
\par We finally make some remarks on
 the flows generated by $\varepsilon_k$,
which, for clarity, we recast to the form (summed on $nm$)
$$ \eqalign{\varepsilon_k &= a_{k0} {\bf H} + a_{k1}
{\hat{\bf H}} + a_{k(mn)}
C_{nm} + {\hat a}_{k(mn)} {\hat C}_{nm} \cr
&= a_{k0} {\bf H} + a_{k1} {\hat{\bf H}} + {\bf Tr}({\tilde a}_k
{\tilde C}) + {\bf Tr} ({\hat{\tilde a}}_k {\hat{\tilde C}}) .\cr}
\eqno(3.17)$$
\par The Poisson bracket $(1.8a)$ then contains a term which is the
$t$-derivative, by $(1.9b)$, but there are additional terms
of general type $(1.10b)$.  In Ref. 10, it is shown that
the terms in $(3.17)$ which contain
${\tilde C},~{\hat{\tilde C}}$ induce transformations on phase
space which are commutators with ${\tilde a}_k,~{\hat{\tilde a}}_k$
in the boson sector, and with ${\tilde a}_k $ in the fermion sector, but
anti-commutator with ${\hat{\tilde a}}_k$ in the fermionic sector.
Hence the elements of the diagonalization transformation
act as connection forms under evolution generated by the effective
mode energy functionals.  Further discussion and application of
these results will be given in Ref. 12.
\bigskip
\noindent
{\bf Acknowledgements}
\smallskip
\par This  work was supported by the Department of Energy under
Grant DE-FG02-90ER40542. One of us (L.P.H.) wishes to thank  C. Piron and
D. Moore for discussions, and for directing us to the
work of Ref. 11.
\bigskip
\noindent
{\bf References}
\smallskip
\item{1.} S.L. Adler, Nuc. Phys. B {\bf 415}(1994) 195.
\item{2.} S.L. Adler, {\it Quaternionic Quantum Mechanics and
Quantum Fields}, Oxford University Press, New York and Oxford, 1995.
\item{3.} E.C.G. Stueckelberg, Helv. Phys. Acta {\bf 33} (1960) 727;
{\bf 34} (1961) 621, 625; {\bf 35} (1962) 673.
\item{4.} D. Finkelstein, J.M. Jauch, S. Shiminovich, and D.
Speiser, J. Math. Phys. {\bf 3} (1962) 207; {\bf 4} (1963) 788.
\item{5.} L.P. Horwitz and L.C. Biedenharn, Ann. Phys. {\bf 157} (1984)
432.
\item{6.} C. Piron, {\it Foundations of Quantum Physics},
W.A. Benjamin, Reading, MA, 1976.
\item{7.} S.L. Adler, G.V. Bhanot and J.D. Weckel, J. Math. Phys.
{\bf 35} (1994) 531.
\item{8.} S.L. Adler and Y.-S. Wu, Phys. Rev. {\bf D49} (1994) 6705.
\item{9.} S.L. Adler and A.C. Millard, ``Generalized Quantum Dynamics
as Pre-Quantum Mechanics,'' Nuc. Phys. {\bf B}, in press.
\item{10.} S.L. Adler and L.P. Horwitz, Institute for Advanced Study
preprint IASSNS-HEP-96/36, ``Microcanonical Ensemble and Algebra
of Conserved Generators for Generalized Quantum Dynamics,''
to be published in Jour. Math. Phys.
\item {11.} R.S. Ingarden, Ann. Inst. Henri Poincar\'e, Vol. VIII, no.1,
(1968)1. See also, R.S. Ingarden and A. Kossakowski,
 Reports on Math. Phys. {\bf 24}(1986) 177, and Roman S. Ingarden,
{\it Termodynamika Statystyczna}, Uniwersytet Mikolaja Kopernica,
Skrypty i Teksty Pomocnicze, Toru\'n 1979.
\item{12.} S.L. Adler and L.P. Horwitz, in preparation.
\item{13.} K. Huang, {\it Statistical Mechanics}, John Wiley and Sons,
New York 1987.

\bye
\end